\documentclass[proof]{WileyASNA-v1}

\articletype{Original Paper}%

\received{11. September 2025}
\revised{12. November 2025}
\accepted{17. November 2025}
\doiheadtext{DOI: \url{https://doi.org/10.1002/asna.70067}}

\begin{document}
\title{Commissioning an Inexpensive Off-the-shelf Spectrograph for Radial-Velocity Studies}
\author[1]{Lukas Stock}
\author[1]{Andreas Schrimpf}

\address[1]{\orgname{Philipps Universität Marburg}, \orgaddress{\state{Hessen}, \country{Germany}}}

\corres{Lukas Stock,  \email{lukas.stock@physik.uni-marburg.de}}

\abstract{We present a way to set up an inexpensive out of the shelf spectrograph at a local observatory. Stability and resolution of the spectrograph are high enough for radial velocity determination of binary stars or determination of stellar characteristics. Even some exoplanets might be detectable via the radial velocity method.}

\maketitle

\section{Introduction} \label{sec:intro}
Telescopes of less than 1-meter aperture are playing an increasingly important role in astronomical investigations. In photometric studies such telescopes are well established for follow-up measurements and extended observational campaigns. Spectroscopic surveys usually are limited in the observation time for single targets and thus demand for continuous follow-up measurements, too. However,
spectroscopic studies with modest-sized telescopes depend on the availability of efficient and easy-to-use spectrographs with sufficient resolution and stability. Spectroscopy in a wider spectral range with single-slit spectrographs is very time-consuming because of the limited spectral range of these instruments. Echelle spectrographs seem to be better suited because of the much wider spectral range of a single exposure and the easily achievable resolving power in the range of $10,000$ to $20,000$.

At the current time to our knowledge there exist only two commercially available echelle spectrometers: BACHES was developed at ESO and the MPE in 2007 \citep{avilla2007}, eShel was designed a bit later by the French company Shelyak \citep{thizy2011}. The BACHES spectrograph is installed, e.g., at an observatory in South Africa \citep{kozlowski2014, kozlowski2016} and a small observatory in Germany \citep{irrgang2016}. The eShel spectrograph is reported to be in operation at several observatories, see ,e.g., \cite{csak2014}, \cite{pribulla2015}, \cite{engel2017}, \cite{gatkine2018} and \cite{rattanasoon2024}.

\cite{eversberg2016} presented extensive tests of both instruments, resulting in a rather low efficiency but a five times higher reliable radial velocity accuracy of about $250\,  \text{m}\, \text{s}^{-1}$ for the Shelyak instrument. \cite{csak2014} and \cite{pribulla2015} claim a short-term stability below $100 \, \text{m}\, \text{s}^{-1}$. The Shelyak instrument is used in recent investigations of non-radial pulsations of a 4-magnitude fast-rotating star  \citep{rattanasoon2024}, monitoring of novae, e.g. \cite{2024ATel16442....1S} and \cite{2018ATel11859....1B}, or a RV-study of a sample of bright non-eclipsing binary candidates \citep{2020MNRAS.497.4884E}.

Our interests are time-domain studies of the radial velocity of hot, massive stars. However, for many of the OB binaries, only a small number of spectra at different epochs are available from LAMOST \citep{liu2024L, yan2022}. We decided to design a completely remote usable instrument with the Shelyak eShel as the main component for usage at a nearby observatory with a 50 cm aperture telescope. In Section \ref{sec:description} the instrument is introduced, in Section \ref{sec:electronics} we describe the electronics and the remote control system. The description of the software for remote control of parameter settings, autofocus adjustment, calibration and object measurements is contained in Section \ref{sec:software}. In Section \ref{sec:Characterization} we present a detailed characterization of the final instrument and Section \ref{sec:usage} describes the scientific value of the instrument.

\section{Description of the spectrograph} \label{sec:description}
The heart of our instrument is an eShel II spectrograph from Shelyak Instruments that was obtained in early 2020. The spectrograph, a ﬁber-fed echelle spectrograph, has three main components: a fiber injection and guiding unit (FIGU) mounted in the telescope's focus, the spectrograph, where the data acquisition camera is attached, and the calibration unit with flat field and calibration light sources. The spectrograph and calibration unit have been placed in a self-designed enclosure for protection and temperature control.

\subsection{Spectrograph} \label{subsec:spectrograph}
The eShel spectrograph itself has been described in detail by \cite{thizy2011}, \cite{pribulla2015} and \cite{eversberg2016}. It was upgraded to eShel II in 2018. The main difference for users is the extension of the wavelength range to 390 --  750 nm. Therefore, we will only briefly discuss the eShel components and focus on our additions.

\begin{figure}[t]
\centering
\includegraphics[width = 0.48 \textwidth]{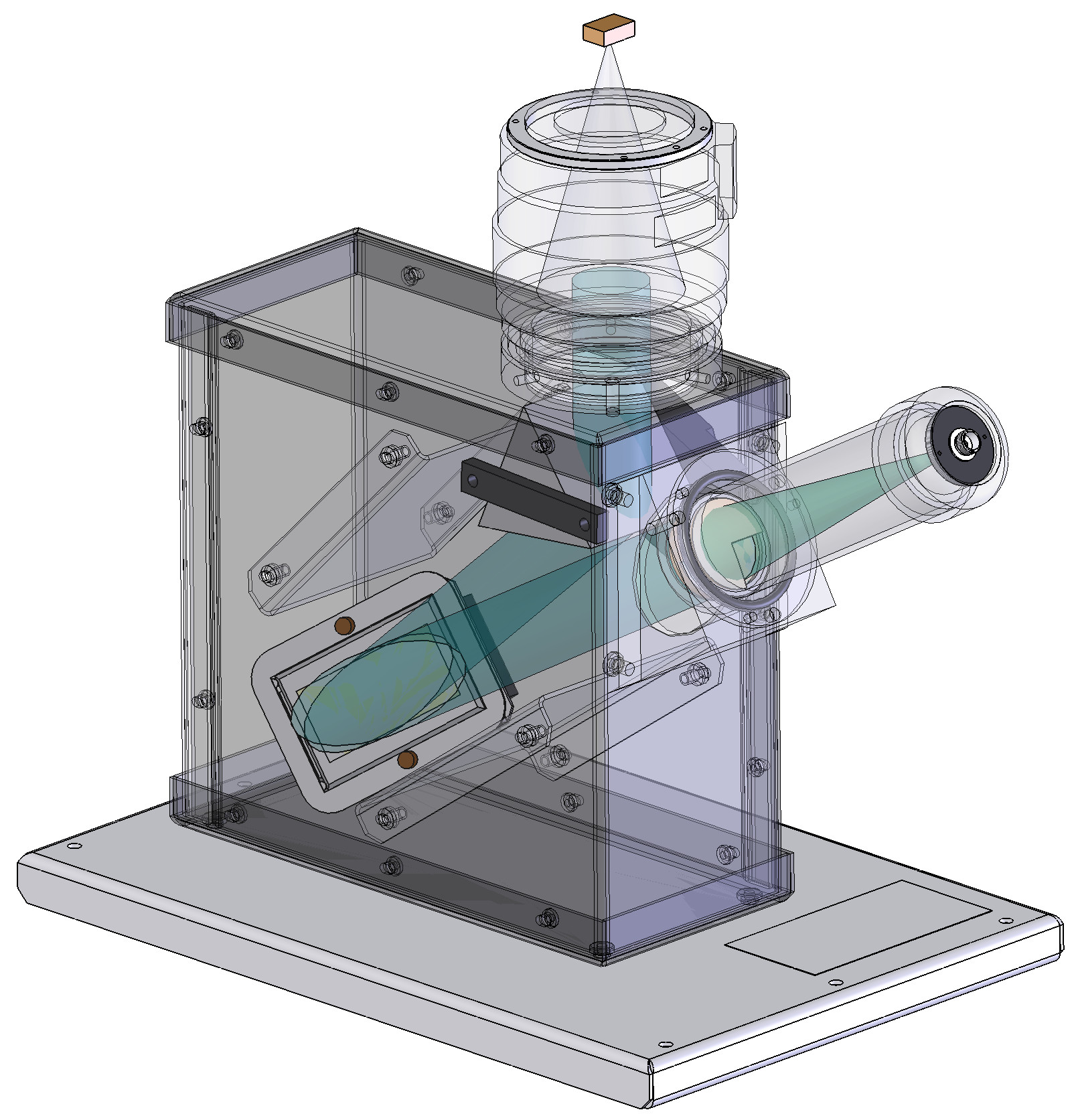}
\caption{Sketch of the Shelyak eShel spectrograph. The stellar light is fed into the spectrograph by an optical fiber on the right side (not shown). The light is then dispersed by a blaze grating, with the echelle orders being separated by a prism. The light beam is then projected onto a camera sensor using a camera lens. Image credit: Shelyak Instruments \label{fig:eshel_sketch}}
\end{figure}

The spectrograph is fed by a 50 $\mu$m fiber with an input $f$-ratio of $f/6$. Inside the fiber injection and guidance unit, the fiber sits behind a small hole in a mirror. To monitor the position of the object of interest on this mirror, a guiding camera can be attached. Fibers can only accept certain $f$-ratios, so the telescope has to match $f/6$ (if necessary by using a focal reducer or a Barlow lens). Shelyak offers a $f/9$ fiber injection and guiding with a 75 $\mu$m hole in the mirror and a built-in reducer to match the 50 $\mu$m fiber. \footnotemark[1]\footnotetext[1]{\tiny \url{https://www.shelyak.com/produit/systeme-complet-eshel/?lang=en}}

Inside the spectrograph, the incoming light is collimated and directed onto an echelle grating. The grating has 79 lines per millimeter and is operating with an incident angle of  $63.45^\circ$ with usable spectral orders from 32 to 53. The spectrum is cross-dispersed by a prism and finally imaged onto the focal plane by a commercial $f$/1.8 Canon lens. Our instrument uses the default $f=85$ mm lens, offered by Shelyak, but lenses with larger focal lengths are available on request. \footnotemark[1]

A Peltier-cooled Atik 383L+ camera is used for data acquisition. It has a monochrome Kodak KAF-8300 CCD chip with 3354x2529 pixels, a pixel size of $5.4 \, \mu \text{m}$ and a pixel scale of about 0.006 - 0.012 nm per pixel in the blue or red part of the spectrum respectively\footnotemark[2]\footnotetext[2]{\tiny \url{https://www.atik-cameras.com/long-exposure-ccd/atik-383l/}}. The backfocus range of the lens is less than 9 mm; the correct adapter has to be chosen carefully. An in-house developed remote-controlled motor-focus system has been added.

The calibration unit holds a ThAr lamp for wavelength calibration, as well as a halogen tungsten lamp and an LED for blaze treatment and flat exposures. A 200  $\mu$m fiber connects the calibration unit with the fiber injection and guiding unit. Different light sources can be turned on and off remotely and a mirror that can be controlled by remote allows switching between sources in the FIGU.

\subsection{Enclosure} \label{subsec:enclosure}
\begin{figure}[t]
\centering
\includegraphics[width = 0.48 \textwidth]{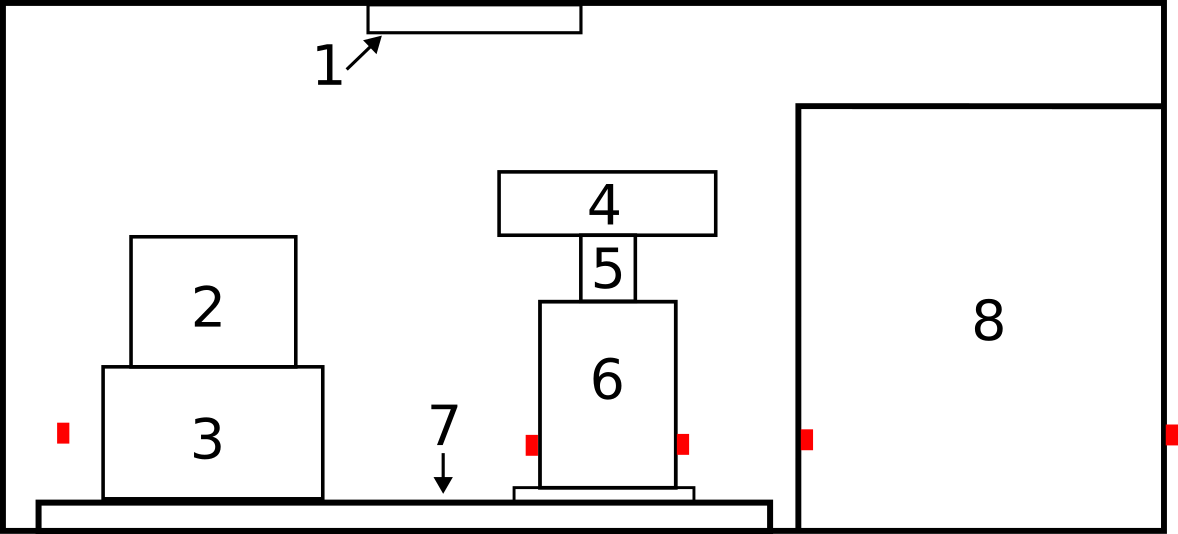}
\caption{Sketch of our enclosure. At the top one of the fans (1) is visible. To the left, the power supply for the ThAr lamp (2) and the calibration unit (3) are displayed. In the middle the spectrograph (6) with camera (4) and lens (5) is shown. Both the control unit and the spectrograph are placed on top of a styrofoam plate (7) to mitigate vibrations. The electronics chamber (8) is displayed on the right. Temperature sensors are highlighted in red. In total, the enclosure measures 76 cm x 56 cm x 61 cm. \label{fig:Box_overview}}
\end{figure}

The only component attached to the telescope is the fiber injection and guiding unit. The two fibers (for observations and calibration) have a length of 20 m each. By selecting cables of suitable length for the remote control of the FIGU and the guiding camera, all other components of the spectrograph can be housed in a shielded box for mechanical protection and reduced stray light (see Fig. \ref{fig:Box_overview}). Besides the eShel components, the box also contains power supplies, temperature sensors, a Raspberry Pi computer and electronics to enable remote control of the entire instrument via LAN. The only connections to the box are the two fibers, a LAN and a power supply cable. There is no need to open the box during observations.

The enclosure is divided into a main and an electronics compartment. The main chamber is completely lined with velours to minimize stray light. This chamber houses the eShel itself, the calibration unit and the camera. 
There are two stacks of equipment in the main compartment. One stack consists of the ThAr lamp power supply on top of the calibration unit. The other stack is the eShel spectrograph with camera, lens and motor focus on top. Both stacks are placed on a polystyrene plate to reduce vibration.

The electronics chamber is separated from the main chamber by a built-in partition and a removable lid. Access to the electronics chamber is only possible through the main chamber and therefore only when the instrument is not in use.

The enclosure also includes slowly rotating fans in the front, the back and the lid of the box. These fans are used to cool the instrument down to the surrounding temperature to avoid thermal stresses on the instrument as far as possible. On the inner side of the fans,  black plates have been mounted to let the air move but block as much stray light as possible. As the enclosure with the spectrograph is located in an open dome next to the telescope, the housing will follow the temperature changes of the night. To log these temperature changes five temperature sensors are mounted in- and outside of the enclosure.

\section{Electronics and remote control} \label{sec:electronics}
\subsection{Remote control}   \label{subsec:remotecontrol}
The eShel spectograph is a fixed setup; there is no need for any direct access or control. The calibration unit comes with a serial remote control interface (RS232) for turning on and off the calibration light sources and switching the mirror in the FIGU. 
The autofocus system for the camera lens (see Section \ref{subsec:autofocus}) we added to the box is remotely controllable and the camera also allows remote access.

To control the electronics a Raspberry Pi is included in the electronics chamber. Camera and calibration unit are directly connected to USB ports of this computer. A USB-serial adapter is used for the RS232 connection. The autofocus electronics as well as the temperature sensors are connected to the GPIO bus of the Raspberry Pi.

An INDI server is running on the Raspberry Pi including drivers from the INDI library\footnotemark[3] \footnotetext[3]{\tiny \url{https://docs.indilib.org/}} and self-made drivers for the hardware. It is connected to the LAN of the observatory.

The Raspberry Pi works as a simple relay station. It takes care of the hardware, for example accessing the camera, autofocus and temperature sensors. No data processing takes place on the Raspberry Pi, all data is routed directly to a remote computer.
This raises the need for a dedicated control software to use the instrument for observations (see Section \ref{sec:software}), but also allows for keeping the Raspberry system as simple as possible. Once the INDI server is running, no maintenance is needed for the Raspberry Pi, making it much easier to handle. 

\subsection{Circuit board}        \label{subsec:circuit_board}
The department's electronic shop designed an electronic circuit board for accessing the additional hardware components we added to the spectrograph. The board contains the connections between the Raspberry Pi and the periphery, namely the stepper motor of the motor focus system and the temperature sensors. The circuit board also serves as a central power distribution point for the fans.

\subsection{Autofocus}        \label{subsec:autofocus}
For enabling readjustments of the focal point, e.g., due to temperature shifts, we designed an autofocus system for the lens on the output of the eShel.

The light from the eShel is focused on the camera chip using a camera lens, in our case an off-the-shelf Canon EF 85mm f/1.8 USM. In combination with a Canon camera the electronics of this lens is powered by the camera and this can be used to automatically focus on objects. However, this is not possible in the usage presented here. Manually setting and adjusting the focus was not an option for a remotely controllable spectrograph, so a motor focus system was designed.

\begin{figure}[t]
\centering
\includegraphics[width = 0.48 \textwidth]{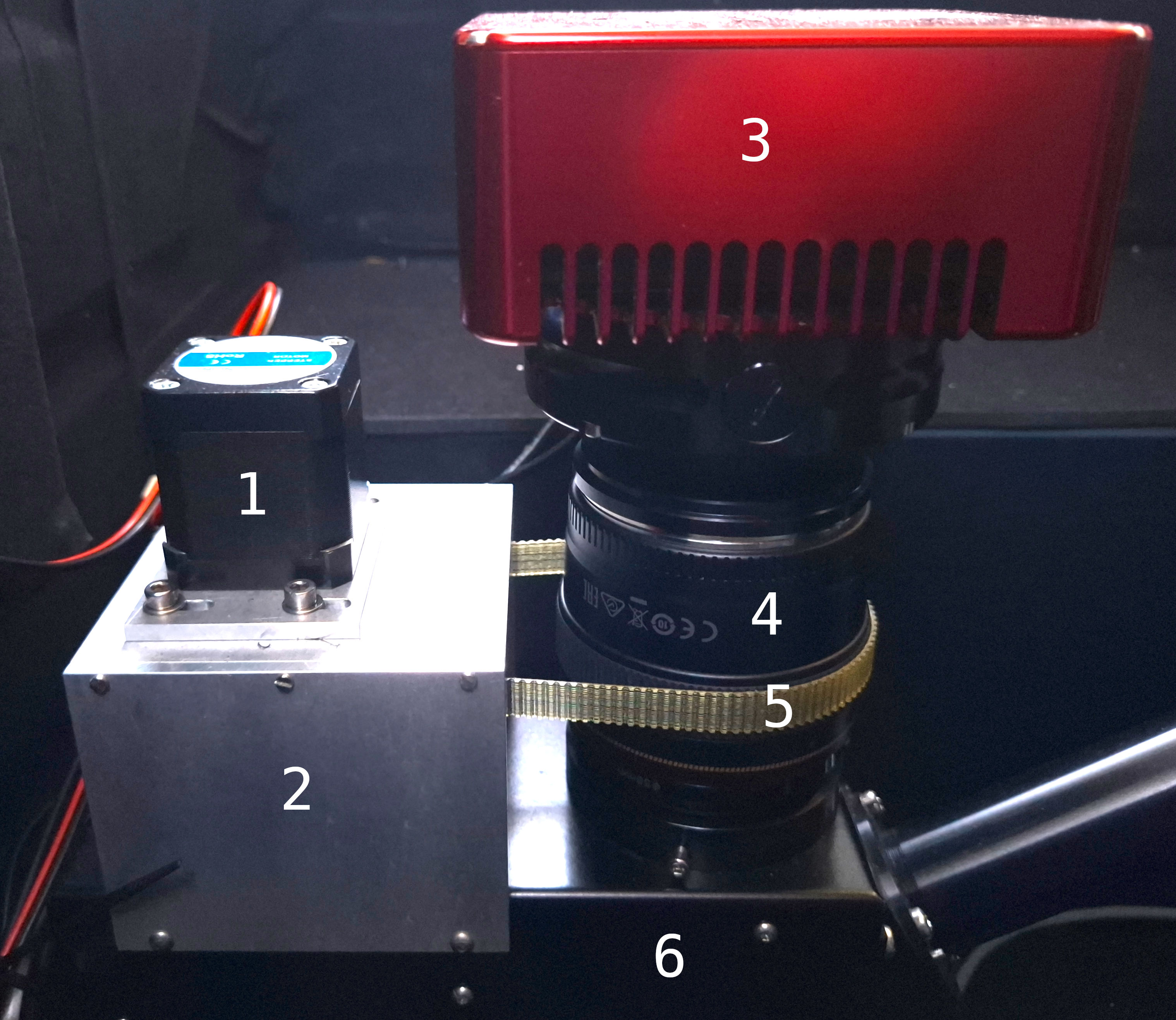}
\caption{The autofocus hardware: On the left side the stepper motor (1) and the attachment of the motor to the spectrograph (2) are visible. This attachment is a U-shaped aluminium plate, made in the department's precision engineering workshop. On the right side the camera (3), the camera lens (4) and the toothed belt between motor and lens (5) can be seen. The spectrograph (6) comes in view below.\label{fig:autofocus} }
\end{figure}

The focus of the lens can be adjusted by turning the focus ring using a toothed belt and a stepper motor. The stepper motor is mounted on top of the eShel, next to the camera lens (see Fig. \ref{fig:autofocus}).

The stepper motor driver is connected to the GPIO pins of the Raspberry Pi via the circuit board. An INDI driver for the stepper motor was developed and installed within the INDI server, so that the stepper motor can be steered with the remote control software (see Section \ref{subsec:software_autofocus}).

We use a NEMA 17 stepper motor with a step width of $1.8 ^\circ$. With this motor we achieve a higher step resolution of about $0.1^\circ$ per step by running the motor in 1/16 microstepping mode. The autofocus routine can move the motor in multiples of this minimum step width.

\section{Remote control software} \label{sec:software}
We have developed a software package that makes controlling the observations remotely as convenient as possible. This software has to handle data from and to the Raspberry Pi. It includes an INDI client, which communicates with the INDI server running on the Raspberry Pi. The software has to control the calibration unit of the eShel, the camera and the motor focus as well as to collect the data from the temperature sensors in the box and the camera. The control software is written in Python 3 and can be used on any Linux PC. 

The software offers a graphical user interface with target, exposure, autofocus and preview tabs to provide easy access to different tasks of the application. The software stores all observational data as FITS files.

\subsection{Target tab} \label{subcec:software_oberview}
In the target tab you can enter the name and/or the star's coordinates, which will then be stored in the header of the final FITS file. This is very helpful, because the remote control software has no connection to the telescope mount itself, so our software has no direct access to information about the target. A built-in query to the SIMBAD database \citep{Wenger2000} ensures the correct name and coordinates of the target.

\subsection{Exposure tab} \label{subsec:software_exposure}
The most important feature of our software package is the ability to plan and execute exposure schedules. One can choose between different exposure modes (all kinds of calibration or light), the exposure time and the number of exposures. It is possible to queue these different types of exposures so that the software can take care of the observations throughout the night. 

Besides the ability to start exposures manually, there is the possibility to start calibrations automatically at the beginning of the observations (mainly Bias, Flat, Orderdef and ThAr frames) and at the end of the observations in the same way as at the beginning, but with dark frames as well. The exposure times of the dark frames correspond to those used for scientific observations. 

The software is handling all parts of the instrument automatically, so there is no need for the observer to ,e.g., manually change the position of the mirror in the FIGU or turn on a calibration lamp. When choosing an observation mode all settings are applied automatically. After finishing the observation queue, all lamps and mirrors will be turned off, so that especially the vulnerable ThAr lamp is not used longer than needed.

\subsection{Autofocus tab} \label{subsec:software_autofocus}
\begin{figure}[t]
	\includegraphics[height=7pc]{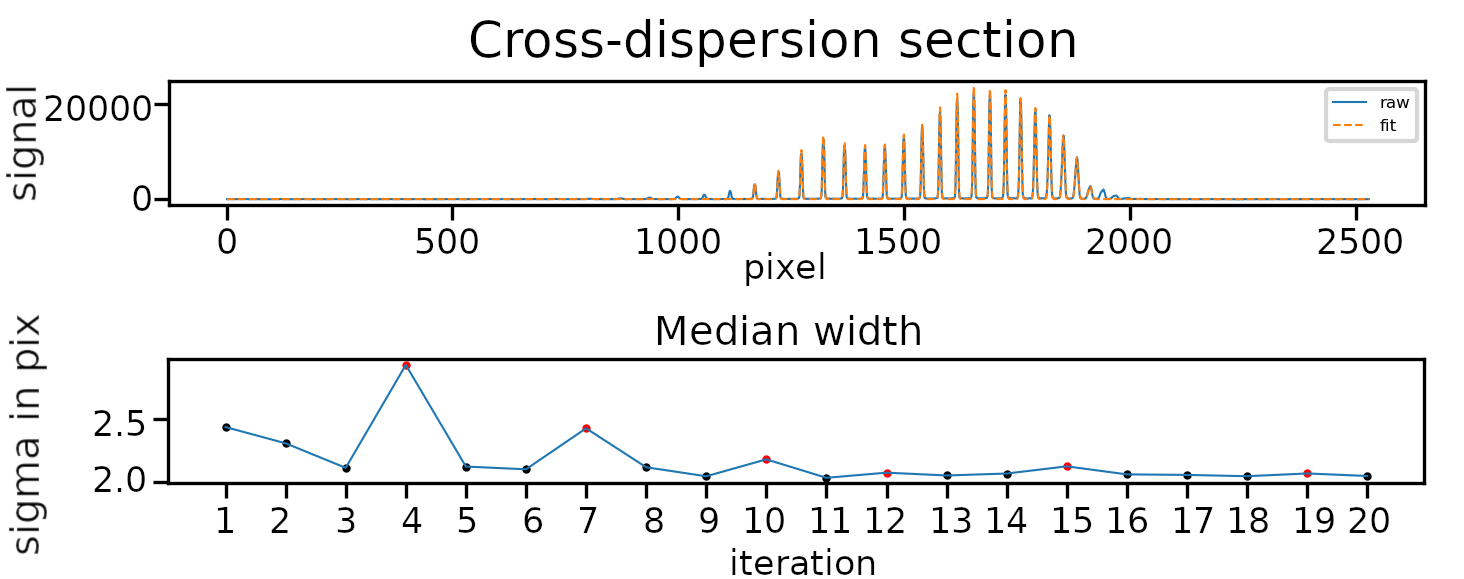}
	\includegraphics[height=7pc]{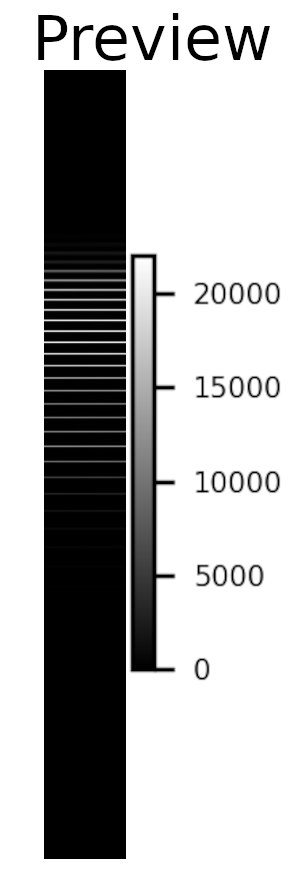}
	\caption{Overview of the autofocus routine. Left: At the top a cross-dispersion cross-section of the flat spectrum can be seen. The single orders are clearly visible and are fitted with a Gaussian, each. The median of the widths of these fits over the subsequent iterations is shown below. Provided the median width is decreasing, the direction of the focus change remains unchanged.  When the median increases the focus direction is reversed and the step width of the autofocus motor decreased. The routine iterates as long as the number of motor steps per iteration is not smaller than 6. Right: Strip of the flat spectrum, where the orders can clearly be seen. Calculating the median of this strip in the dispersion direction generates the 1D-spectrum used for the autofocus routine. \label{fig:software_autofocus}}
\end{figure}

In Section \ref{subsec:autofocus}, we presented the hardware of the motor-focus system that we attached to the optics of the spectrograph. However, to use the hardware to automatically set the best focus possible, we had to design an autofocus routine.  

The routine uses images of the cross-dispersed orders of the flat spectrum. With optimized focus, the widths of the tracks of these orders on the CCD are minimal. To determine the widths, we extract a 5-pixel-wide strip from the center of the CCD. We take the median over those 5 pixels, identify all orders via a search for maxima and fit all orders separately with a Gaussian each. The median of the widths of all these Gaussians is a measurement of the focus that needs to be minimized. 

The optimum focus is determined by a control loop. In each loop the median of the widths of the flat spectrum orders is calculated. Starting with 50 steps of the focus-stepper motor the focus is changed. If the median of the widths has increased, the direction of the stepper motor rotations is reversed. If it decreases, we repeat this step until we overshoot the focus and the median of the widths increases again.
This time, the direction of the stepper motor is reversed and the number of steps is reduced by a factor of 1.5. When changing focus direction one has to take into account the backlash, which was determined to be 35 steps in our system. We decrease the step width until we reach a minimum of 6 steps per focus change. At a smaller number of motor steps we cannot detect any focus changes safely. Once the focus gets worse again at 6 steps, we simply move backward these 6 steps and the backlash and stop there. 

In the autofocus page, one can check the cross dispersion of the spectrum in the strip, the current fit to the single maxima, as well as the median Gaussian width over different iterations (see Fig. \ref{fig:software_autofocus}). This allows the observer to check autofocus behavior, although the software operates completely autonomously.

The focus position will change slightly during the night due to temperature shifts. We do not correct for those shifts, as this could change the spectral order position relative to the calibration frames taken at the beginning of the observation.

\subsection{Preview tab} \label{subsec:software_preview} 
The remote control software offers the possibility to preview an extracted spectrum of the last captured image. This allows the observer to verify that the spectrum appears as expected and that no errors, e.g., a defective calibration lamp or false pointing of the telescope, occurred. To create the extracted spectra, we use a fast extraction method of our data reduction software (see Section \ref{subsec:usage_datareduction}) that is included in the control software. The preview should not be used for scientific evaluations, as the data reduction is kept very simple to provide just a quick preview.

\section{Characterization of the spectrograph} \label{sec:Characterization}
\subsection{Resolution} \label{subsec:spec_Resolution}
\begin{figure}[b]
	\centerline{\includegraphics[width=0.48 \textwidth]{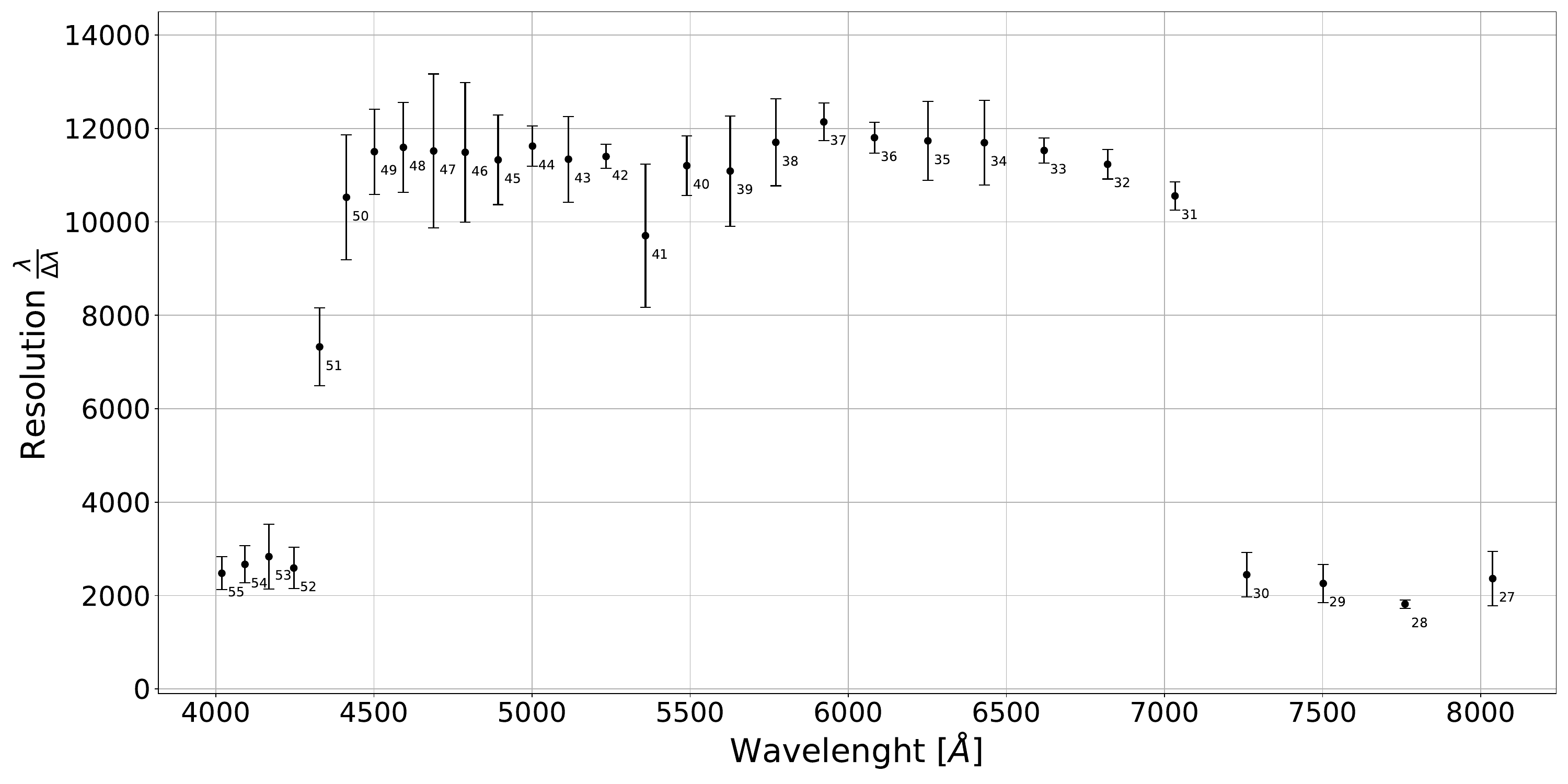}}
	\caption{Resolution of the eShel II spectrograph in different orders or wavelengths, respectively. The resolution drops near the edges of the spectral range due to optical errors in the system, which increase towards the edges of the image sensor. \label{fig:Resolution}}
\end{figure}

The Shelyak eShel II is designed to provide a resolution of $\geq 10 \, 000$ over the whole spectral range\footnotemark[4]\footnotetext[4]{\tiny \url{https://www.shelyak.com/produit/pf0011-spectroscope-eshel/?lang=en}}. We can confirm that this resolution is achieved and even exceeded in the medium wavelength range as shown in Fig. \ref{fig:Resolution}. However, the resolution of the spectrograph drops significantly at the edges of the spectral range. This is mainly due to an amplification of the optical errors of the system, especially of the used lens. Correcting these optical errors should lead to a significant improvement in resolution in the edges. Alternatively, using a lens with higher optical quality should also result in fewer optical errors in the low and high orders. When changing the lens, especially when using a lens with a different focal length, one has to make sure the size of the camera sensor still fits the output or also change the camera. \\
\begin{figure}[t]
	\centerline{\includegraphics[width=0.48 \textwidth]{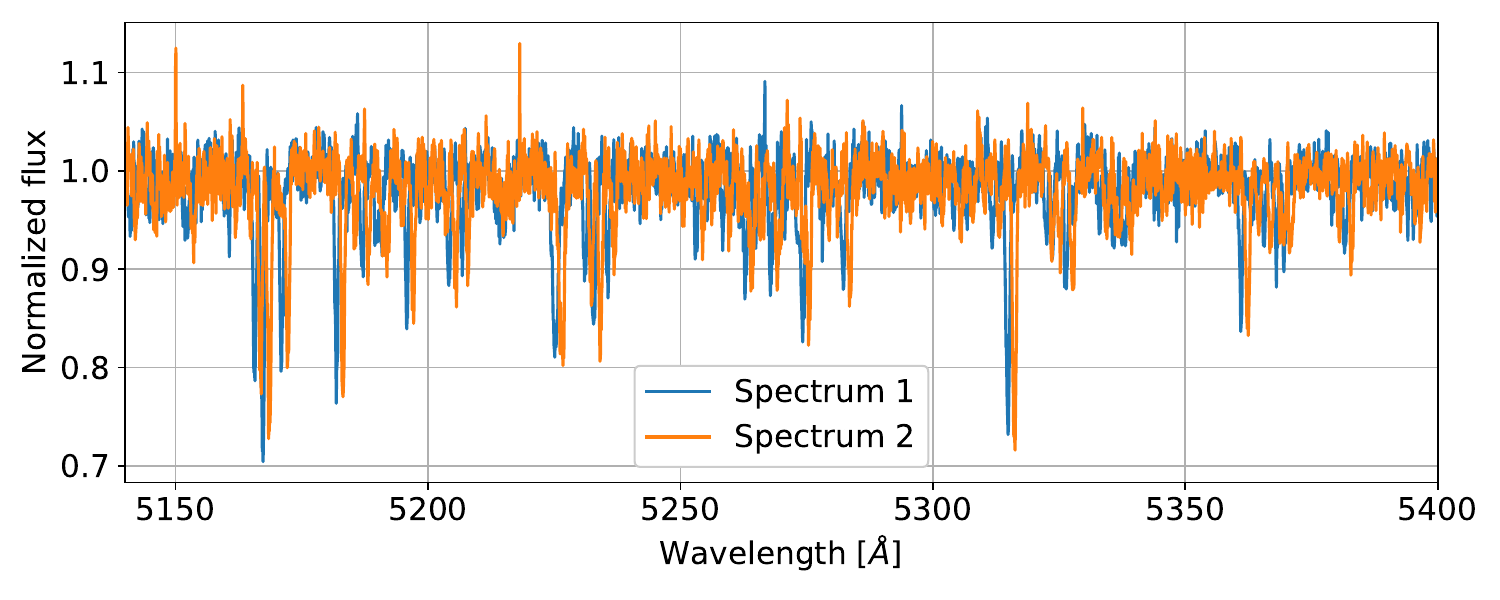}}
	\caption{Detail of the spectrum of 21 Cas. Two spectra of the same star at different times are shown and the spectral shift between those spectra is clearly visible.  The resolution of the instrument is more than sufficient to resolve the spectral features to allow an accurate radial velocity determination.  The RV shift between the two spectra is $\Delta RV = 73.31 \pm 0.76$ km s$^{-1}$.\label{fig:21_Cas_spec}} 
\end{figure}

However, it must be mentioned that we can measure orders that are not officially listed in the documentation of the spectrograph. Shelyak mentions a spectral range of about $3900$ to $7000 \, \textup{~\AA}$ or orders 32 to 53 respectively\footnotemark[5]\footnotetext[5]{\tiny \url{https://www.shelyak.com/produit/systeme-complet-eshel/?lang=en}. \\ \hspace*{0.58cm} The values listed on the website still refer to the eShel I version of the spectrograph.}. We, on the other hand, were able to measure the orders 27 up to 55, extending the spectral range up to $3900$ to $8200 \, \textup{~\AA}$. We do not see an increase in the spectral range in the blue part of the spectrum because the orders there overlap very strongly and are therefore only very slightly shifted in relation to each other. This being said, we can operate in regions where the spectrograph is not optimized, but these ranges are still usable, albeit with reduced resolution and efficiency. 

The resolution of the spectrograph is sufficient for different science fields such as, e.g., radial velocity determination or spectral classification. This will be explained in greater detail in Section \ref{subsec:usage_science}. As an example, Fig. \ref{fig:21_Cas_spec} shows the spectrum of the binary system 21 Cas, which will be discussed in further detail in Section \ref{subsubsec:21Cas}. The shift caused by the different radial velocities is clearly visible, which underscores the usefulness of the spectrograph.

\subsection{Stability} \label{subsec:spec_Stability}
\begin{figure}[t]
\centering
\includegraphics[width = 0.48 \textwidth]{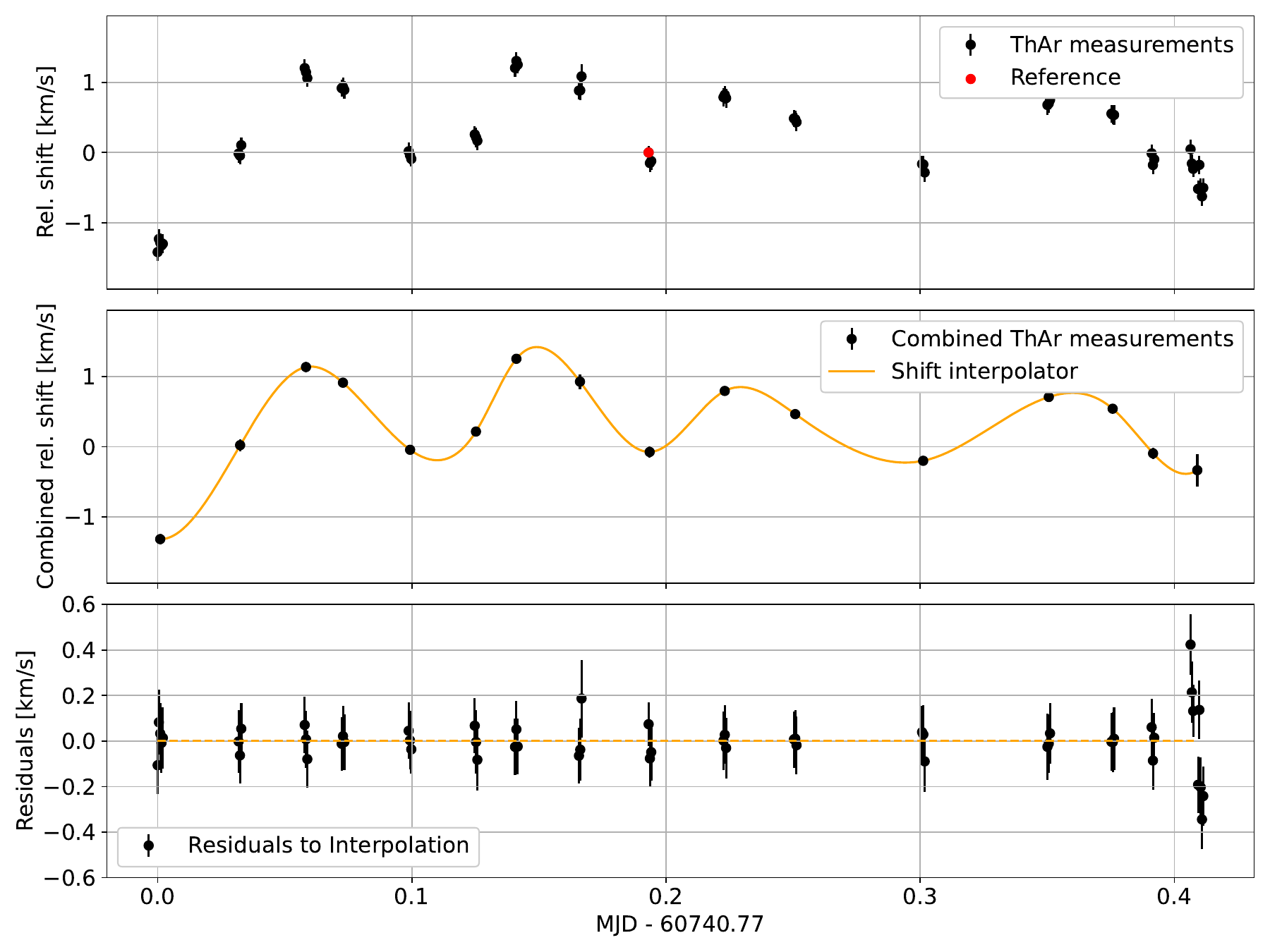}
\caption{Relative shifts of ThAr calibration frames over the course of one night. The best wavelength solution was used as a reference. 
Top: Shift of every taken ThAr calibration frame to the reference. The reference solution is marked in red. At each observation time three ThAr frames are taken directly one after the other and there should be nearly no difference between those. This can be seen in the image, as the three consecutive ThAr frames show nearly no shift relative to each other. The accuracy of an individual wavelength solution is about $150 \, \text{m s}^{-1}$ here. 
Center: Related frames are grouped to reduce the error. Between those calibration points a cubic spline is used to interpolate the shift for the light frames. 
Bottom: Residuals of the initial ThAr measurements to the interpolation, which is created using the grouped measurements. Taking into account individual errors, the measurements scatter around the interpolation in the range of approx. $250 \, \text{m s}^{-1}$. This value is a conservative estimate of the error, as no data exists for the science measurements between the calibration exposures. \label{fig:shift_interpolation}}
\end{figure}

To convert the raw spectra from the intrinsic pixel-based representation to a wavelength-based representation, we need a conversion formula between those two. This formula is the result of the wavelength calibration using the ThAr frames and we will call it the wavelength solution. Because the ThAr frames differ from one another, the wavelength solutions belonging to those frames will also differ. The stability of the instrument is characterized by the size of the difference between the wavelength solutions.

The stability of the instrument depends on several factors, such as temperature and pressure, but also on effects like vibrations or internal processes in the ThAr lamp, which is used for wavelength calibration. As we use the instrument in an open dome we cannot control the ambient temperature or the air pressure nor can we avoid any vibrations caused by moving the telescope or the dome. We try to minimize the influence of the latter by placing the eShel on top of a polystyrene plate (as described in Section \ref{subsec:enclosure}). All effects can affect the accuracy of the instrument on small time scales, especially within one exposure, but also cause a systematic drift. 

We let both the dome (so the local environment of the instrument) and the instrument itself cool down to the ambient temperature before starting the observations. By means of the fans described in Section \ref{subsec:enclosure}, the initial temperature gradient between instrument and environment is minimized.

\begin{figure}
\centering
\includegraphics[width = 0.48 \textwidth]{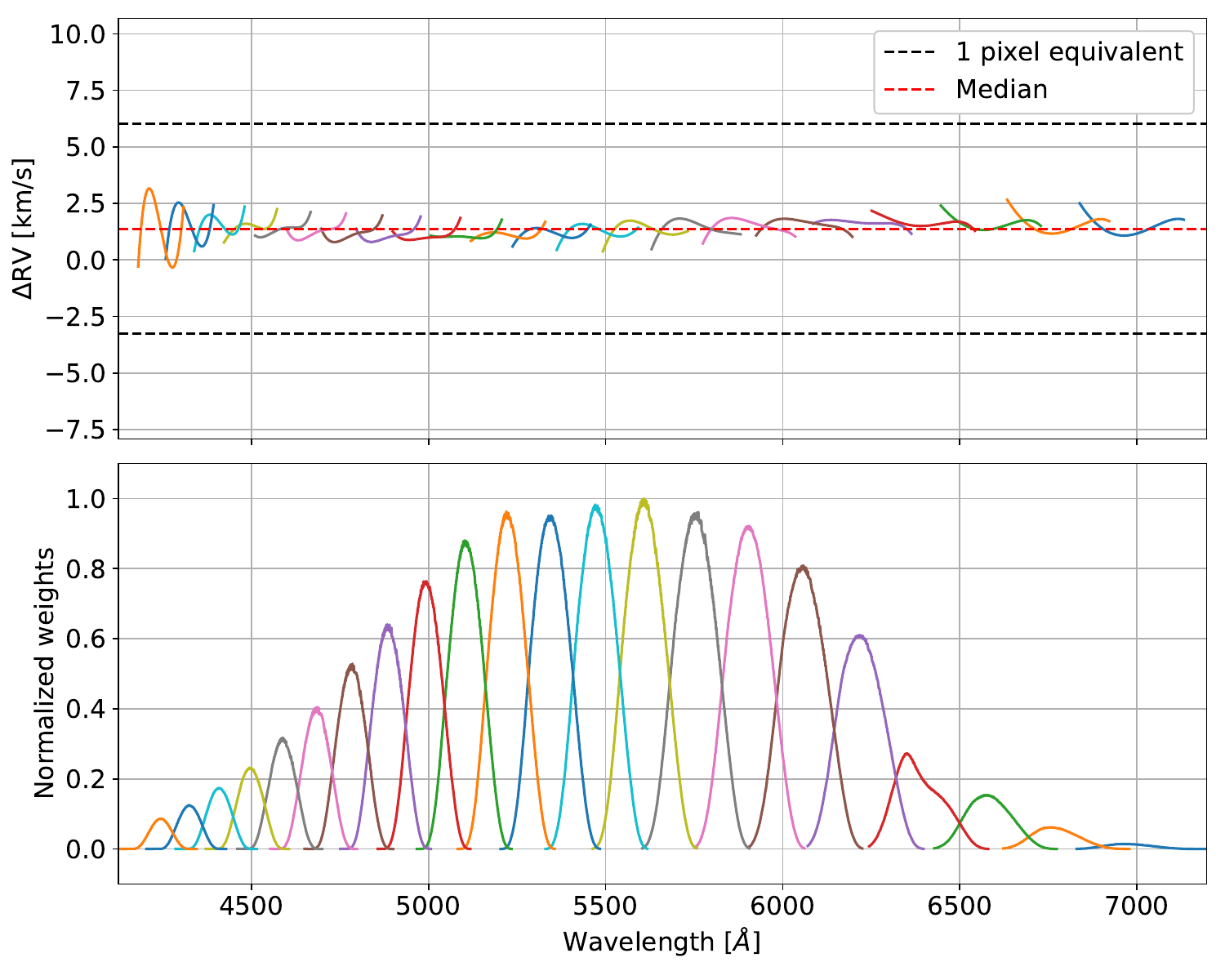}
\caption{Top: Residuals between a chosen reference wavelength solution and another wavelength solution from the same night. Only residuals with a weight of at least 0.05, relative to the corresponding order, are shown. The individual orders of the instrument are shown in different colors. Bottom: Weights for each pixel for individual orders. The weights correspond to the SNR of each pixel. It is clear that the boundaries of each order are insignificant, as there is almost no signal. 
\label{fig:Wavelength_res}
}
\end{figure}

The ambient temperature will drop overnight. This will also affect the temperature of the instrument during operation, leading to small changes in the optical system caused by thermal expansion. Usually all of these changes are described as a shift in radial velocity compared to a reference wavelength solution of the calibration process (in detail described in a future paper)
\begin{align*}
\lambda = \left(1 + \frac{v}{c} \right) \cdot \lambda_\text{ref}
\end{align*}
where $\lambda$ is the wavelength at each pixel, $v$ is the radial velocity shift between the two solutions and $c$ is the speed of light in vacuum \citep{Brahm2017}.

In our case, the thermal changes and vibrations cause a drift of the wavelength solution of about $2 \, \text{km  s}^{-1}$. We cannot avoid those changes, but we can monitor the drift overnight by periodically taking ThAr calibration frames and interpolating the drift (see Fig. \ref{fig:shift_interpolation}). This allows us to reduce the total error of the calibration to about $250 \, \text{m s}^{-1}$. 

When taking a closer look at the wavelength residuals between two wavelength solutions, one recognizes that the median shift is indeed almost constant across different spectral orders and wavelengths, as can be seen in Fig. \ref{fig:Wavelength_res}. This is not the case for all wavelengths individually, however. The interpolation of the wavelength solutions using Chebyshev polynomials leads to oscillations, which can be much larger than the accuracy of the wavelength solution. The reason for these oscillations lies in the lack of calibration lines in the border areas of the orders, as there is almost no signal due to vignetting of the instrument. The wavelength solution is still extrapolated to these pixels, leading to errors. However, these errors are insignificant, as there is almost no science signal at these pixels and the errors are mostly smaller than the one pixel accuracy, as can be seen in Fig. \ref{fig:Wavelength_res}. The median of each order also matches nicely with the total median of the residuals, so that there is no wavelength dependency of the shift.

\subsection{Example measurements} \label{subsec:examples}
\subsubsection{Measurement method} \label{subsubsec:measurement_method}
The target star's radial velocity was determined in two different ways, depending on the physical parameters of the system. When only one component of the system is visible in the spectrum, e.g., because the star is orbited by an exoplanet or a much fainter companion, the radial velocity is derived using the cross-correlation function (CCF) method. The spectrum with the best SNR is used as a template. In this way the relative RV to the template can be determined very accurately. It is not possible to calculate absolute radial velocities, but this is not mandatory for determining the orbit. 

If two or more components of the system are visible in the spectrum, the spectrum consists of two components that shift against each other. Therefore, the spectrum of the star itself is not a good template. Instead, we use synthetic spectra from the POLLUX database \citep{Palacios2010}. We also do not use the CCF method in this case, as the broadening function (BF) method delivers better results here \citep{Rucinski2002}.

\subsubsection{21 Cas} \label{subsubsec:21Cas}
\begin{figure}[b]
\centerline{\includegraphics[width=0.48 \textwidth]{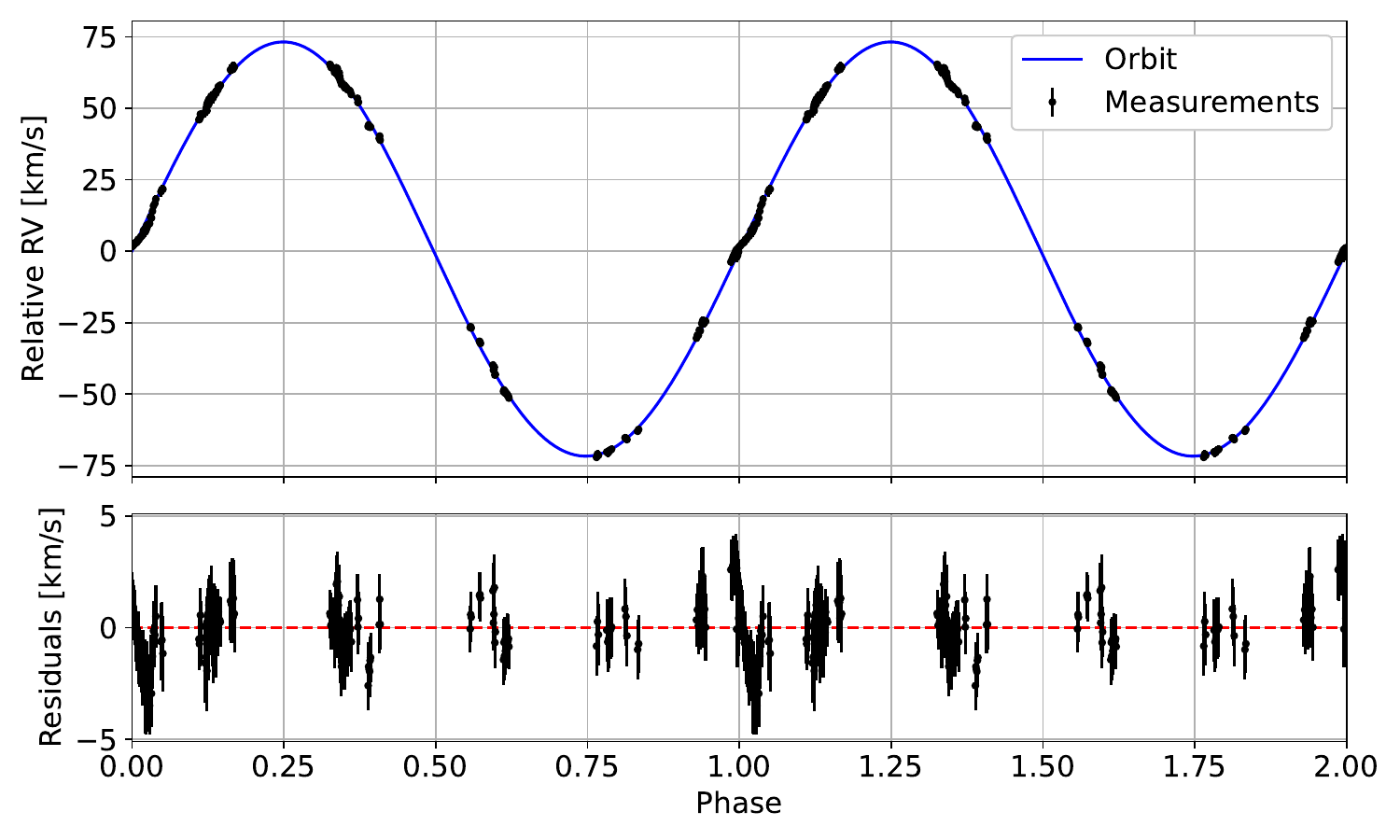}}
	\caption{Top: Phasefolded RV plot of 21 Cas. The RV curve is plotted twice to highlight the periodicity. 
	 Bottom: Residuals to best orbit fit. 
	 \label{fig:21Cas_RVs}}
\end{figure}

\begin{table}[b]
\begin{tabular}{|l|c|c|}
\hline
 & Own measurements & Pavlovski et al. (2014) \\
\hline 
$K$ [km s$^{-1}$] & $72.46 \, \pm \, 0.15\phantom{0}$ & $73.05 \, \pm 0.19$\\
\hline
$P$ [days]      & $4.46711 \pm 0.00006$ & $4.4672235 \text{ (fixed)}$ \\
\hline
$\epsilon$      & $0.011 \, \pm \, 0.003$ & $0$ (fixed) \\
\hline
\end{tabular}

\caption{Comparison of our derived parameters for the orbit of 21 Cas with the ones from \protect\cite{Pavlovski2014}. $K$ is the velocity amplitude, $P$ the orbital period and $\epsilon$ the orbital eccentricity. \label{tab:Params_21Cas}} 
\end{table}
21 Cas (or YZ Cassiopeiae) is a known eclipsing binary system about 340 light-years away from Earth. It has an apparent V-magnitude of about 5.7 \citep{Hoeg2000}. The primary star is an A-type star of about 2.3 solar masses. It is orbited by a companion star of about 1.3 solar masses in approximately 4.46 days \citep{Lacy1981, Pavlovski2014}. 

21 Cas was observed in eleven nights between August 2023 and March 2025. This allows us to fit its orbit with high precision using the "exoplanet" package \citep{ForemanMackey2021}. Our results and a comparison to those of \cite{Pavlovski2014} are listed in Table \ref{tab:Params_21Cas}. In Fig. \ref{fig:21Cas_RVs} the radial velocity plot and the orbital fit are shown. The comparison between orbital fit and the observations proves the accuracy of our instrument.

\subsection{GK Cep} \label{subsubsec:GKCep}
\begin{figure}[b]
	\centerline{\includegraphics[width=0.48 \textwidth]{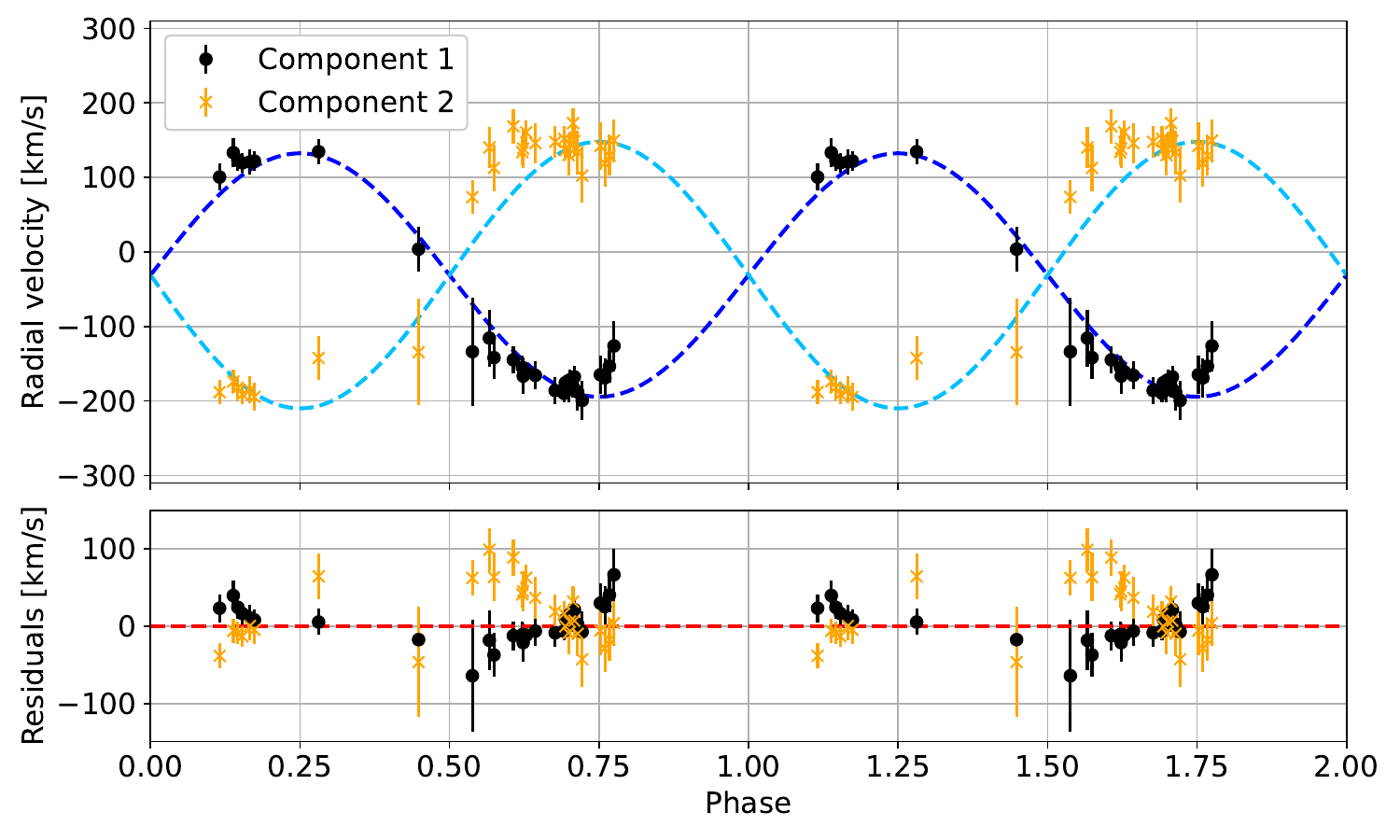}}
	\caption{Top: Phasefolded RV plot of GK Cep. Both components of the system are displayed, the main component with black circles and the companion with orange crosses. The shown orbits are derived from the parameters given by \protect\cite{Pribulla2009}. The RV curve is plotted twice to highlight the periodicity. Some measurements were rejected due to spectral line blending, especially close to phases  0 and 0.5.
	 Bottom: Residuals to known orbits. \label{fig:GKCep_RVs}}
\end{figure}
GK Cep is a long-period W UMa star system with a period of about 0.936 days \citep{Derman1992}. The system has a V-magnitude of 7.0 \citep{Hoeg2000}. Both components of the system are type A2V stars \citep{Derman1992} and as such they show nearly no spectral features besides very broad Balmer lines, resulting in a challenging radial velocity determination. Since both components of the system have approximately the same brightness, both components are visible in the spectrum. The broadening function (see Section \ref{subsubsec:measurement_method}) was used to determine the radial velocity for both components. 

GK Cep was observed in 15 nights between May 2023 and March 2025. Due to its relatively low magnitude, the SNR of this star's spectra is only about 30, which results in some loss of RV accuracy. For this star, the typical RV accuracy is 20 km s$^{-1}$, as can be seen in the Fig. \ref{fig:GKCep_RVs}. Nevertheless, our results are in good agreement with the results of \cite{Pribulla2009}.

\subsection{$\tau$ Boo} \label{subsubsec:TauBoo}
\begin{figure}[t]
	\centerline{\includegraphics[width=0.48 \textwidth]{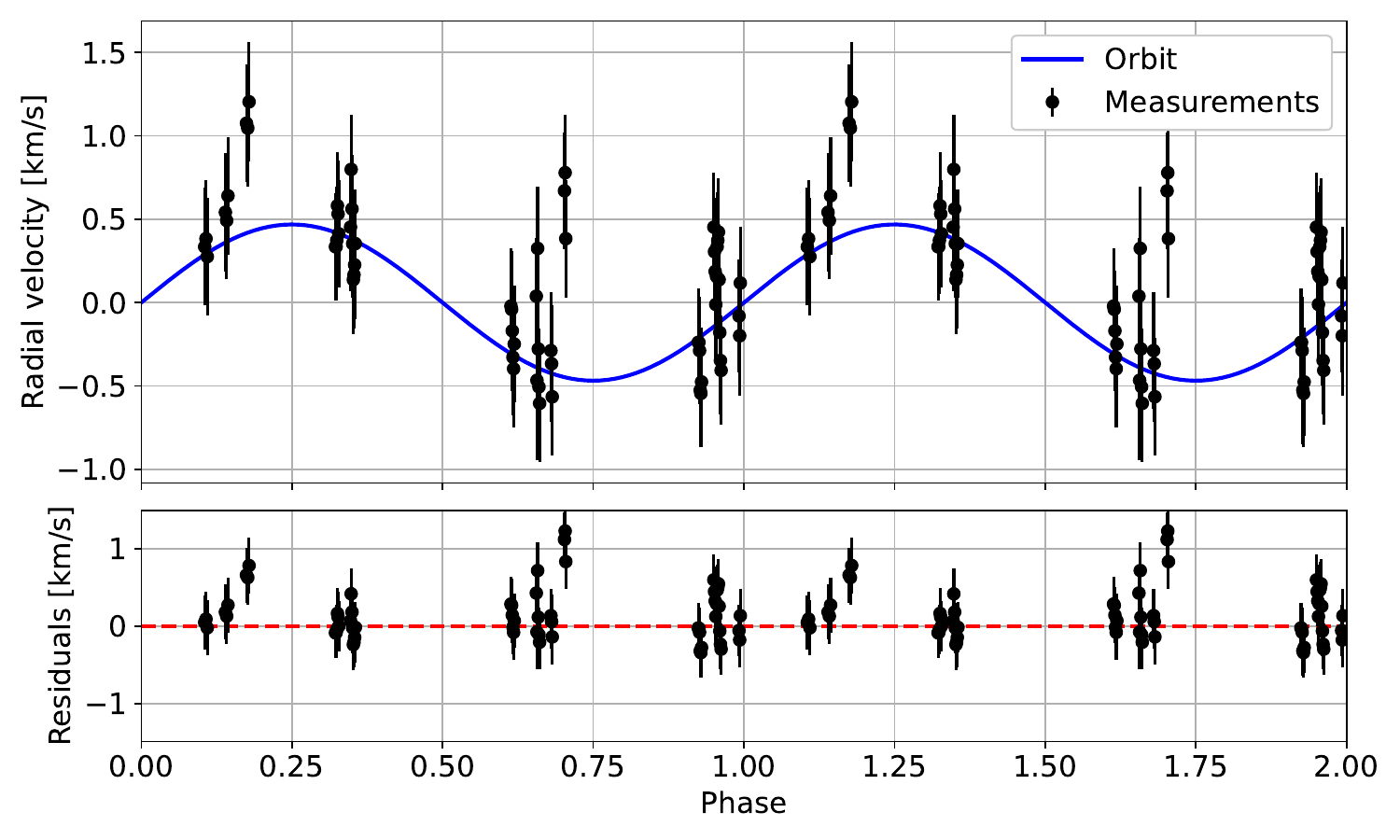}}
	\caption{Top: Phasefolded RV plot of $\tau$ Boo. The shown orbit is derived from the parameters given by \protect\cite{Justesen2019}. The system velocity was adjusted to better match our data. The RV curve is plotted twice to highlight the periodicity. 
	 Bottom: Residuals to known orbit. \label{fig:TauBoo_RVs}}
\end{figure}
$\tau$ Boo (or Tau Boötis) is a bright (V$_\text{mag} = 4.49$ \citep{Belle2009}) binary star system. The main component ($\tau$ Boo A) is a F7V-type star and hosts an exoplanet ($\tau$ Boo b), which is known since 1997 \citep{Butler1997}. The exoplanet orbits its host star in about $P \, = \, 3.3125$ days. The radial velocity amplitude is $K \, = \, 468$ m s$^{-1}$ \citep{Justesen2019}. This target was chosen, because the radial velocity signal is in the same order of magnitude as the stability of our instrument. By studying this target we can demonstrate the abilities of the instrument.

$\tau$ Boo was observed in five nights in February and March 2025. Fig. \ref{fig:TauBoo_RVs} shows the previously known orbit of $\tau$ Boo A and our measurements. While some outliners are present, the orbit is still clearly visible. This is especially remarkable as the radial velocity variation is less than the instrumental drift  described in Section \ref{subsec:spec_Stability}.

\section{Usage} \label{sec:usage}
\subsection{Data reduction} \label{subsec:usage_datareduction}
For data reduction we use a software based on the package CERES \citep{Brahm2017}. We updated the software to Python 3 and expanded its functionality. 

The software uses a new extraction routine, including recognition of and extraction along tilted spectral lines, as necessary with this instrument. Also the extraction pipeline supports a robust wavelength calibration method using overlaps of neighboring orders, which was first described in \cite{Brandt2020}. Many pipelines require a list containing each ThAr emission line and, in particular, its position on the detector chip. This list is difficult to create for an instrument that has not yet been characterized. In our approach, we only need a general list of the wavelengths of the ThAr emission lines, but we do not need to know the exact position of these lines on the detector chip.

All new features of the updated software package will be presented in a future paper.

\subsection{Science field} \label{subsec:usage_science}
Our instrument provides the advantage of being usable at any time the time schedule requires. This makes it ideal for studying rapidly changing astronomical sources, such as close binary star systems. Characterizing binary star systems needs many spectra to provide all information about the system such as orbital period, inclination or masses of the single components. Usually one relies on archive data, where there can be a long time span without any data from that system. To study close systems with periods of hours to days, one needs a very high cadence. Our instrument can provide this. 

On the other hand, the instrument can also be used for educational purposes. It provides an intuitive entrance into spectroscopy and allows pupils and university students to get used to spectrographs. Besides studying radial velocities one could also characterize known stars or study exotic stars such as ,e.g., Wolf-Rayet stars. 

The limiting magnitude of the instrument depends on the scientific field. For radial velocity studies we estimate a SNR of 10 to be the absolute minimum, whereby a loss of accuracy is already noticeable here. With our telescope and a 10 min exposure time this corresponds to a star with V$_\text{mag} = 9.3$. If it is sufficient to just detect spectral features, we estimate that a SNR of 2 would be enough. With the same assumptions as above this corresponds to a limiting magnitude of V$_\text{mag} = 12.8$.

\subsection{Best practice} \label{subsec:usage_bestpractice}
We describe a best practice on how to use the instrument during an observation. Before starting the observation, one has to ensure that dome, telescope and instrument are adapted to the ambient temperature to avoid thermal stresses during observation as much as possible. The first step is to focus the spectrum on the camera sensor with the camera lens. This can be done, as in our case, with an autofocus motor, or manually. It is also recommended to create Bias, Flat, Orderdef and ThAr exposures before starting with the scientific observations, as these calibration frames are essential for data reduction. In our case we create 5x0.2s Bias, 3x10s Flat, 3x10s Orderdef and 5x30s ThAr frames. Taking multiple frames allows elimination of contamination by cosmic rays via taking the median. 

If RV analysis is the goal of the observation, we recommend to take  ThAr frames after moving the telescope (e.g. because of a different target star, tracking neglected) to monitor the stability of the instrument when telescope and dome move. For the same reason we recommend to also do three ThAr frames before moving the telescope the next time. Even when the target stays the same overnight, we recommend doing ThAr exposures at least every hour. However, this can take a non-negligible amount of time, so if no high wavelength stability is needed (e.g., because the goal is to determine the spectral type of the target star) it is sufficient to just take ThAr frames at the beginning and the end of the observation.

\section*{Acknowledgements}
The author thanks MARA, the MArburg University Research Academy, for financial support. The author thanks Shelyak for providing additional images for illustration. This publication made use of the python packages Lmfit \citep{Newville2014}, Matplotlib \citep{Hunter2007}, Numpy \citep{Harris2020} and Scipy \citep{Virtanen2020}.

\section*{Conflicts of Interest}
The authors declare no conflicts of interest.

\newpage

\bibliography{eShel}{}

\end{document}